\documentclass[12pt,psf,epsf]{JHEP3}
\usepackage{euscript,amsmath,amssymb,epsfig,latexsym,amsfonts}

\usepackage{graphicx}

\newcommand{\be}{\begin{equation}}
\newcommand{\ee}{\end{equation}}
\newcommand{\bea}{\begin{eqnarray}}
\newcommand{\eea}{\end{eqnarray}}
\newcommand{\nn}{\nonumber}

\title{Note on the Unruh Effect}
\author{I.Ya. Aref'eva and I.V. Volovich\\
\small{\it Steklov Mathematical Institute, RAS, Moscow}}
\date {\today}
\maketitle

\abstract{
It was suggested by Unruh that a uniformly accelerated detector in vacuum
would perceive a noise with a thermal distribution.

We obtain  a  representation of  solutions of the wave equation in
two dimensions suitable for the Rindler regions.
The representation  includes the dependence on a parameter.
The Unruh  field corresponds to a singular limit of the representation.

}

\begin{document}

\section{Introduction}

It was suggested that a uniformly accelerated detector in vacuum in
the Minkowski space would perceive a thermal bath of particles
\cite{Unruh}. This phenomenon is known as the Unruh effect, see
\cite{Takagi, DeWitt, Crispino} for review.
 It is supposed   that the Unruh quantum
 field in the Rindler region is the ordinary quantum field but only
expended in terms of another modes related with the original ones by
means of  the Bogolyubov transformation.

We first consider the classical field.
We obtain a representation of the two-dimensional massless field
in Minkowski space by using the Mellin transform.
There is a parameter $\lambda$ in our representation.
If $\lambda >0$ then we have the ordinary field.
The formal limit $\lambda \rightarrow 0$ corresponds to the Unruh field.
In fact one can perform the limit $
\lambda \rightarrow 0$  for  classical solutions with the special boundary condition,
but not for arbitrary   solutions.  Therefore
the classical version of the Unruh field is not equal
  to the ordinary  field.

Our representation after the standard quantization
 produces a set of modes that also depend on $\lambda$.
 From these modes
in the singular limit $\lambda\rightarrow 0$
one can get  the thermal vacuum distribution.

\section{Classical Field}

\subsection{Wave equation}
Let us discuss first solutions of the classical  two-dimensional
wave equation
 \be (\partial ^2_t-\partial ^2_x)\Phi(x,t)=0\ee
We will consider smooth solutions $\Phi(x,t)\in C^{\infty}({\bf
R}^2)$ belonging to to the space $S({\bf R}^1)$ of fast decreasing
functions  over the variable $x$. By using the Fourier transform one
can write the solution in the standard form

\bea \label{Phi-a} \Phi(x,t)&=&\frac{1}{\sqrt{4\pi}}\int
_{-\infty}^\infty
[a(k)e^{i(kx-t|k|)}+a^*(k)e^{-i(kx-t|k|)}]\frac{dk}{\sqrt{|k|}}\eea
where $a(k)$ is a fast decreasing function. This formula can be
written in the form
 \be\label{Phix}
\Phi(x,t)=F(u)+G(v),\,\,\,\,\,u=x-t,\,\,\,v=x+t\ee where\be
F(u)=\frac{1}{\sqrt{4\pi}}\int _{0}^\infty
[a(k)e^{iku}+a^*(k)e^{-iku}]\frac{dk}{\sqrt{k}}\label{F-A}\ee
\be\label{Gv} G(v)=\frac{1}{\sqrt{4\pi}}\int
_{0}^\infty[a(-k)e^{-ikv}+a^*(-k)e^{ikv}]\frac{dk}{\sqrt{k}}\ee Note
that $F,G\in S({\bf R}^1)$. Writing $G(v)=\theta (v)G_+(v)+\theta (-v)G_-(-v)$ and
$F(u)=\theta (u)F_+(u)+\theta (-u)F_-(-u)$ one gets the representation of the
solution of Goursat problem for the wave equation in the $R, L, F$ and $P$ regions
(in L and R region one interchanges time and space coordinates).

\subsection{Mellin transform}

The plane wave in terms of the Rindler coordinate $\xi = \log v$
reads $e^{i\omega\xi}=v^{i\omega}$ that corresponds to the Mellin
transform.

For a function $G\in S({\bf R}^1)$ the Mellin transform
of its restriction $G_+$ to the positive half-line $v>0$ is defined by
\be\label{Mellin} \widetilde G_+ \left( s \right) =\int _{0}^{\infty
}G_+ \left( v \right){v}^{s-1} { dv},\,\,\,\,\Re s>0.
\ee
$\widetilde G_+
\left( s \right)$ admits an analytic continuation to a meromorphic
function in the whole complex plane  with simple poles at
$s=0,-1,...\,$. For small $s$ one
has $\widetilde G \left( s \right)=G_+(0)/s+...$

 The inverse Mellin transform is given by
\be\label{invMellin} G_+(v)= \frac{v^{-\lambda}}{2\pi }\int
_{-\infty}^{\infty}  \widetilde
G_+(\lambda+i\omega)v^{-i\omega}d\omega, \,\,\,\,\,\,v>0 \ee where
$\lambda$ is an arbitrary positive number, we will assume
$0<\lambda<1/2$.

For real valued functions $G_+$ by using (\ref{Gv}), (\ref{Mellin}),
(\ref{invMellin}) and the relation

\be \int_0^{\infty}v^{s-1}e^{-ikv}dv=k^{-s}e^{-\frac{i\pi
s}{2}}\Gamma (s) \ee
one gets the representation
\be\label{G}
G_+(v)=\frac{v^{-\lambda}}{2\sqrt{\pi} }\int _{0}^{\infty}[
B_+(\omega,\lambda) v^{-i\omega}+ B_+^*(\omega,\lambda)
v^{i\omega}]\frac{d\omega}{\sqrt{\omega}},\,\,v>0,\lambda >0.
\ee
Here \be\label{B_+}
 B_+(\omega,\lambda)=\Gamma(\lambda+i\omega)\frac{\sqrt{\omega}}
 {2{\pi}}\int _{0}^\infty
dk \,k^{-i\omega-\frac12-\lambda}\left[a(-k)e^{\frac{\pi \,\omega}2
} e^{-\frac{i\pi \,\lambda}2 }+a^*(-k)\,e^{-\frac{\pi \,\omega}2 }
e^{\frac{i\pi \,\lambda}2 }\right]
\ee
We can take the limit $\lambda \rightarrow 0$ in (\ref{B_+}) to get
\be\label{B_+0}
 B_+(\omega, 0)=\Gamma(i\omega)\frac{\sqrt{\omega}}
 {2{\pi}}\int _{0}^\infty
dk \,k^{-i\omega-\frac12}\left[a(-k)e^{\frac{\pi \,\omega}2 }
+a^*(-k)\,e^{-\frac{\pi \,\omega}2 } \right]
\ee
This   corresponds to the Unruh-DeWitt modes \cite{DeWitt}.
However we cannot just  substitute (\ref{B_+0}) into the formula
(\ref{G}) for the inverse Mellin transform due to the
singularity $\Gamma (i\omega)\sim 1/i\omega$ as $\omega \rightarrow
0$. The singularity in the inverse Mellin transform is related with  the  boundary
condition $G_+(0)=0$.

The importance of the boundary condition in the Unruh problem is stressed
in \cite{Narozhny}.

  If $\lambda >0$ then we have the ordinary classical field.
Analogously to (\ref{G}) one can write a Mellin representation for
the field (\ref{Phix}) for the functions $G_{\pm}(v)$ and $F_{\pm}(u)$.

\section{Quantum field}
Quantization of the field $\Phi(x,t)$ (\ref{Phi-a}) can be performed
by postulating the canonical commutation relations: \be\label{com}
[a(k),a^*(k')]=\delta(k-k') \ee To cure the infrared divergence we
introduce the cut-off $\kappa >0$.

From (\ref{B_+}) and (\ref{com}) we obtain
\be\label{B+B-p} [ B_+(\omega,\lambda),B_+^*(\omega',\lambda)]
\\ =
\Gamma(\lambda+i\omega)\Gamma(\lambda-i\omega')\frac{\sqrt{\omega\omega'}}
{2\pi^2}\sinh
(\frac{\pi (\omega+\omega '}2)\int _{\kappa}^\infty dk
\,k^{-i(\omega-\omega')-1-2\lambda}. \ee
The integral is \be\label{int} \int _{\kappa}^\infty dk
\,k^{-1-2\lambda-i(\omega-\omega')}=\frac{\kappa^{-2\lambda-i(\omega-\omega')}}
{2\lambda+i(\omega-\omega')}\ee In the limit $\lambda\rightarrow 0$
and $\kappa \rightarrow 0$ the integral goes to the $2\pi$ times the
$\delta$-function and one can write
\be \nn[ B_+(\omega, 0),B_+^*(\omega', 0)]=|\Gamma (i\omega)|^2\frac{\omega}{2\pi^2}
\sinh(\pi\omega)2\pi\delta(\omega-\omega')=
\delta(\omega-\omega').
\ee
However as it was discussed earlier we should keep $\lambda>0$ for
the ordinary field.

The formula (\ref{B_+0}) provides a Bogolyubov transformation for
the Unruh field. For the ordinary field we have the formula
(\ref{B_+}).

For the vacuum ($ a(k)|0>=0$ ) expectation value we obtain
\bea &\,& <0|\, B_+^*(\omega,\lambda) B_+(\omega,\lambda)\,|0>=
\Gamma(\lambda-i\omega)\Gamma(\lambda+i\omega)\frac{\omega}{4\pi^2}
e^{-\pi \omega }\frac{\kappa^{-2\lambda}}{2\lambda}\label{inter}
\eea
Since \be \Gamma(-i\omega)\Gamma(i\omega)\omega e^{-\pi \omega }=
\frac{2\pi}{e^{2\pi \omega}-1}\ee  the
formula (\ref{inter}) in the limit
$\lambda \rightarrow 0$ (one ignores the divergent constant)
describes the thermal distribution with the temperature
$T=2\pi$.

\section {Conclusions}
In this note  the representation of the solutions of
 the classical two-dimensional massless field is obtained.
 The representation  includes the dependence
  on the parameter $\lambda >0$.
It is shown that after quantization in the singular limit $\lambda\rightarrow 0$
one can get the thermal vacuum distribution.

In quantum field theory and in quantum optics various quantum noises
are occurred if one uses an appropriate limiting procedure such as the
stochastic limit, see \cite{ALV}. It is important to note that the limiting
procedure can be performed only  for functionals of  special form.
In this note
the thermal vacuum noise  in the limit $\lambda\rightarrow 0$ for special functionals
is  obtained.

\section{Acknowledgments} The  work is partially supported by
RFFI 11-01-00894-a (I.A.) and RFBR 11-01-00828-a (I.V).

\end{document}